\documentclass[11pt]{article}\topmargin=-0.7cm \textheight 222mm \textwidth 15.8cm\oddsidemargin=4mm
\usepackage{amssymb}
\newcommand{\comm}[1]{}
\def\short{\comm}

\def\noxxx{\comm}

\usepackage{color}
\newcommand{\rd}[1]{\color{red}#1\color{black}} 
\def\rd{}
  
   \csname @addtoreset\endcsname{equation}{section}
\newtheorem{theorem}{Theorem}
\newtheorem{definition}{Definition}
\newtheorem{remark}{Remark}

\def\e{\varepsilon}

\def\defi{\stackrel{{\scriptscriptstyle \Delta}}{=}}

\def\a{\alpha}

\def\o{\omega}

\def\F{{\cal F}}
\def\w{\widehat}
\def\Ind{{\,\rm Ind\,}}
\def\Ind{{\mathbb{I}}}

\def\Var{{\rm Var\,}}
\def\Re{{\rm Re\,}}

\def\R{{\bf R}}
\def\E{{\bf E}}

\def\H{{\cal H}}

\def\h{h}
\def\L{L}

\def\b{\beta}
\def\s{\delta}
\def\g{\gamma}
\def\C{{\bf C}}

\def\X{{\cal X}}
\def\t{\theta}
\def\oo{\bar}
\def\s{\sigma}
\def\D{{\Delta}}

\def\U{{\cal U}}

\def\L{{\cal L}}

\def\h{h}

\newcommand{\be}{\begin{equation}}
\newcommand{\ee}{\end{equation}}
\newcommand{\bd}{\begin{displaymath}}
\newcommand{\ed}{\end{displaymath}}
\newcommand{\ba}{\begin{array}{ll}}
\newcommand{\ea}{\end{array}}
\newcommand{\baa}{\begin{eqnarray}}
\newcommand{\eaa}{\end{eqnarray}}
\newcommand{\baaa}{\begin{eqnarray*}}
\newcommand{\eaaa}{\end{eqnarray*}}
\font\sm=cmr10
\def\GS{{\scriptscriptstyle GS}}

\def\oo{\bar}

\def\a{\alpha}
\def\D{{\cal D}}

\def\HH{\mathbb{H}}
 
\def\ew{\left(i\o\right)}
\def\sinc{{\rm sinc}}
\date{ Submitted: May 19,  2020. Revised: October 8, 2021}
\title{Limited memory predictors based on polynomial approximation of periodic exponents
}
\author{Nikolai Dokuchaev\\
\sm Zhejiang University/University of Illinois at Urbana-Champaign Institute, \\
\sm  Zhejiang University, Haining, Zhejiang Province, China 314400}
 
\begin{document}
 \def\short{\comm}
\def\brea{}
\def\breakk{}
\def\break{}
\def\BRR{}
\def\breakm{\nonumber\\  }\def\BR{}\def\BRR{}
\def\breacm{}
\maketitle
\noxxx{ \let\thefootnote\relax\footnote{Corresponding author. 
 Email: Dokuchaev@intl.zju.edu.cn and ndokuch@me.com}}

\begin{abstract} The paper presents transfer functions for limited memory time-invariant linear integral predictors for continuous time  processes such that the corresponding predicting kernels have bounded support.
It is shown that processes with exponentially
decaying Fourier transforms  are predictable with these predictors in some weak sense, meaning that   convolution integrals over the future times can be approximated by causal convolutions over past times.  For a given predicting horizon,
the predictors  are based on polynomial approximation of a  periodic exponent (complex sinusoid)  in a weighted 
$L_2$-space.  
\par  
\end{abstract}
\section{Introduction}
 We study pathwise predictability and predictors of
continuous time processes in deterministic setting and in the
framework of the frequency analysis. It is well known that certain
restrictions on frequency distribution can ensure additional
opportunities for prediction and interpolation of the processes;
see, e.g.,  Slepian (1978),  Knab
(1979), Papoulis (1985), Marvasti (1986), Vaidyanathan (1987), Lyman
{\it et al} (2000, 2001) and the bibliography therein. 
\index{\citet{K}-\citet{W}}
These works considered predictability of band-limited processes;
the predictors were non-robust with respect to
small noise in high frequencies; see, e.g., the discussion in
 Higgins (1996), Chapter 17.

We consider  some special  linear weak predictability: 
instead of predictability of the original processes, we
study predictability of sets of anticausal convolution integrals
based on linear  time-invariant integral predictors.  This version of
predictability was introduced in Dokuchaev (2008)
\index{\cite{D08}} for  band-limited and high-frequency
processes. In \index{\cite{D21}} Dokuchaev (2021),  the problem was considered  for processes single point spectrum degeneracy.    In Dokuchaev (2021), \index{ \cite{D10},}  the problem was considered  for processes with exponentially decaying Fourier transforms.  
\comm{ In \cite{D17},  a similar predictability  was considered  for processes with a single point spectrum degeneracy.}   In these works, integral predictors with kernels  featuring unlimited support were derived.  Respectively, the predicting  algorithms based
 on these predictors    would require unlimited history  of observations for the underlying processes.   

Following the setting from Dokuchaev (2010), \comm{ \cite{D10},} the present paper considers processes  with exponentially decaying Fourier 
transforms.   It is known that these processes are analytic and therefore allow 
a unique extension from any open interval. In particular, an arbitrarily accurate prediction can be achieved  
via Taylor series expansions wth sufficiently  small steps and sufficiently high order of the Taylor polynomials. However,
 this  would require calculations of large number of derivatives for the underlying processes 
which is impractical. The result of Dokuchaev (2010) \index{\cite{D10}} allowed to use "universal" integral type predictors instead of calculating derivatives of the underlying process. 
The goal of the present paper is to develop integral predictors with  limited memory, i.e. such that the  corresponding convolution  kernels have bounded support, and the corresponding predicting algorithm  requires history 
of observations on some finite time interval.   The setting with limited memory  predictors was first suggested in Lyman {\it et al} (2000)  and Lyman and Edmonson (2001) \index{\cite{L,Ly}}
for stochastic stationary  band-limited  processes.  In Lyman {\it et al} (2000), \index{\cite{L}}, an existence result for the predictors but the predictors are not derived.   In Lyman and Edmonson (2001) \index{\cite{Ly},}   the predictors are obtained for the case of a known preselected spectral density.

In the present paper, some principally   new    predictors are
obtained  via polynomials approximating a periodic exponent $e^{i\o T}$ in exponentially weighted 
$L_2$-spaces, where $\o\in\R$, and where  $T>0$ is a preselected prediction horizon. These predictors allow a compact explicit representation in the time domain
given by equation (\ref{TD}) below. In addition, these predictors
allow explicit representation  in the frequency domain via their transfer
functions given by equation (\ref{FD}) below. The  choice of the predictors is independent on the
spectral characteristics of  input processes.

The paper is organized in the following manner. In Section
\ref{secDef}, we formulate the definitions and background facts related to
the linear weak predictability. In Section
\ref{secM}, we formulate
 the main theorems on predictability and predictors (Theorem \ref{ThM} and Theorem \ref{Th2}).
 In Section \ref{secPol}, we discuss possible choices of approximating polynomials.  In Section \ref{secRob}, we discuss the
robustness of the predictors.
Section \ref{secProof} contains the proofs. Finally, in Section \ref{secCon}, we discuss  our results.

\section{Problem setting and definitions}\label{secDef}
Let $x(t)$ be a currently observable continuous time process,
$t\in\R$. The goal is to estimate, at current times $t$, the values
$y(t)=\int_{t}^{t+T} h(t-s)x(s)ds$, using historical values of the observable process
$x(s)|_{s\le t}$. Here $h(\cdot)$ is a given
kernel, and $T>0$ is a given prediction horizon. \par
 We consider  linear predictors in the form $\w y(t)=\int_{t-\tau}^t\w
h(t-s)x(s)ds$, where $\w h(\cdot)$ is a kernel that has to be found, and where $\tau>0$.
We call $\w k$ a predictor or predicting kernel.
\par
To describe admissible classes of $h$ and $\w h$, we need some notations and definitions.
\par
 Let  $\R^+\defi[0,+\infty)$, $\C^+\defi\{z\in\C:\
\Re z> 0\}$, $\C^-\defi\{z\in\C:\
\Re z< 0\}$, $i=\sqrt{-1}$.
\par
For $p\in[1,+\infty]$, we denote by $L_p(\R,\R)$ and $L_p(\R;\C)$  the usual $L_p$-spaces
of functions $x:\R\to\R$ and $x:\R\to \C$ respectively.
\par
For $x\in  L_p(\R,\C)$, $p=1,2$, we denote by $X=\F x$ the function
defined on $i\R$ as the Fourier transform of $x$; $$X(i\o)=(\F
x)(i\o)= \int_{-\infty}^{\infty}e^{-i\o t}x(t)dt,\quad \o\in\R.$$ If
$x\in L_2(\R,\C)$, then $X$ is defined as an element of $L_2(i\R,\C)$, i.e.,  $X(i\cdot)\in L_2(\R,\C)$.
\par
For $x\in L_p(\R,\C)$, $p=1,2$, such that $x(t)=0$ for $t<0$, we denote by
$\L x$  the Laplace transform \baa\label{Up} X(z)=(\L
x)(z)\defi\int_{0}^{\infty}e^{-z t}x(t)dt, \quad z\in\C^+. \eaa
In this case, $X|_{i\R}=\F x$.
\par
Let  $\HH^r$ be the Hardy space of holomorphic functions $U(z)$ on $\C^+$  with finite norm
$\|U\|_{\HH^r}=\sup_{s>0}\|U(s+i\cdot)\|_{L_r(\R,\C)}$, $r\in[1,+\infty]$; see, e.g., Duren (1970)).
\index{\cite{Du}}

Let  $\HH^r_-$ be the Hardy space of holomorphic functions  $U(z)$ on $\C^-$  with finite norm
$\|h\|_{\HH^r_-}=\sup_{s>0}\|U(-s+i\cdot)\|_{L_r(\R,\C)}$, $r\in[1,+\infty]$.

\begin{definition}
For $\t\in [0,+\infty)$ and  $T\in (0,+\infty)$,  we denote by $\H(-T,\t)$ the set of functions $h:\R\to\R$
such that $h(t)=0$ for $t\notin [-T,\t]$ and such that $h\in
C^\infty(\R)$.
\end{definition}

It can be shown that the traces of functions from $\H(-T,\t)$ are everywhere dense in $L_2(-T,\t)$: let $\varkappa_\e(t)$ be defined as
 $\varkappa_\e (t)=\e^{-1}\varkappa_1(t/\e )$,
where $\varkappa_1(t)$ is the
so-called Sobolev kernel defined as $\varkappa_1(t)=\kappa^{-1}\exp(t^2(t^2-1)^{-1})\Ind_{|t|<1}$,
where $\kappa=\int_{-1}^1\exp(t^2(t^2-1)^{-1})dt$.

 Let $\oo h\in L_2(\R)$ be a function vanishing outside $(-T,\t)$. Then,
for any $\e>0$, functions $h_\e$, defined
as the convolutions
\baaa
h_\e(t)=\int_{-\infty}^{\infty}\varkappa_\e (t-s)\Ind_{[-T+\e,\t-\e]}(s)\oo h(s)ds,\label{filter}
\eaaa
 belong to $\H(-T,\t)$ and approximate $\oo h$ in $ L_2(\R)$.
\begin{definition} For $\tau>0$, let $\w\H(0,\tau)$ be the class of functions $\w h:\R\to\C$ such that $\w h\in L_2(\R;\C)$,  
$\w h(t)=0$ for $t\notin [0,\tau]$  and such that $\w H(\cdot)=\L\w h\in \HH^2\cap
\HH^\infty$. 
\end{definition}
\par
\begin{definition}\label{def1}
Let  $\oo\X$ be a class of processes $x$ from $L_2(\R;\C)\cup
L_1(\R;\C)$.  Let $T>0$, $\t\ge 0$, and $\tau>0$, be given.
\begin{itemize}
\item[(i)]
 We say that the class $\oo\X$ is linearly weakly $(\tau,\t)$-predictable 
 with the prediction horizon $T$ if, for any
$h(\cdot)\in\H(-\t,T)$, there exists a sequence $\{\w
h_{d}(\cdot)\}_{d=1}^{+\infty}=\{\w
h_{d}(\cdot,\oo\X,k)\}_{d=1}^{+\infty}\subset \w\H(0,\tau)$ such that $$
\sup_{t\in\R}|y(t)-\w y_{d}(t)|\to 0\quad \hbox{as}\quad
d\to+\infty\quad\forall x\in\X, $$ where \baaa &&y(t)\defi
\int_t^{t+T}h(t-s)x(s)ds,\qquad\breakk \w y_{d}(t)\defi \int^t_{t-\tau}\w
h_{d}(t-s)x(s)ds.\label{predict} \eaaa The process $\w y_{d}(t)$
is the prediction of the process $y(t)$.
\item[(ii)] Let the set $\F(\oo\X)\defi \{X=\F x,\quad x\in\oo \X\}$  be provided with a norm $\|\cdot\|$.
 We say that the class $\oo\X$ is  linearly weakly $(\tau,\t)$-predictable 
  with the prediction horizon $T$ uniformly with
respect to these norm $\|\cdot\|$, if, for any $h(\cdot)\in\H(-\t,T)$,
there exists a sequence $\{\w h_{d}(\cdot)\} =\{\w h_{d}
(\cdot,\X,h,\|\cdot\|)\}\subset \w\H(0,\tau)$ such that \baaa &&\sup_{t\in\R}|y(t)-\w
y_{d}(t)|\to 0\quad\breakk\hbox{uniformly in} \quad
\{x\in\oo\X:\ \|X\|\le 1,\quad X=\F x\}. \eaaa Here  $y(\cdot)$  and $\w
y_{d}(\cdot)$ are defined in part (i) of this definition.
\end{itemize}
\end{definition}

\begin{remark} 
 We  include  the case where $\t>0$, because the choice of $\t=0$
would allow only  $h$ vanishing at zero. For these kernels, the  values of $x$ at the nearest future 
 times are not covered by the prediction. 
\comm{
We  do not exclude the case where $\t<0$, because the case  $\t=0$
would restrict the choice of  admissible $h$  by the smooth kernels vanishing at zero. 
This would  exclude possibilities of the short term prediction of $x$. }
\end{remark}
\section{The main result} 
\label{secM}
For $r>0$, let  $L_{2,r}(\R;\C)$ be the Hilbert space of processes 
$u:\R\to\C$ with the norm 
 \baaa \|u\|_{L_{2,r}(\R,\C)}=\Bigl(\int_{-\infty}^{+\infty}e^{r|\o|}|u(\omega)|^2 d\o\Bigr)^{1/2}.\eaaa

Let $\X(r)$ be the  set of 
processes $x\in L_2(\R;\C)$ such that $\|X( i \cdot) \|_{L_{2,r}(\R;\C)} <+\infty$ for $X=\F x$.

Let  $\U(r)$   be a class of processes
$x(\cdot)\in \X(r)$ such that $\|X( i \cdot) \|_{L_{2,r}(\R;\C)}\le 1$. 
\vspace{0.4cm}  
\begin{theorem}\label{ThM}  For any $r>0$, $T>0$,  $\t\ge 0$, and $\tau>0$, the following holds.  
\begin{itemize}\item[(i)] The class $\X(r)$ is $(T+\t,\t)$-predictable
in the weak sense with the prediction horizon $T$.
\item[(ii)] The class $\U(r)$ is linearly weakly
$(T+\t,\t)$-predictable  with the prediction horizon $T$
uniformly with respect to the norm $\|\cdot\|_{L_{2,r}(\R)}$.
\end{itemize}
\end{theorem}\vspace{0.4cm}  
\begin{remark}
It can be seen that if $h\in\H(-\t,T)$ 
and $x \in\X(r)$, then $y\in \X(r)$, where $y$ is such as described in Definition \ref{def1}. 
It follows  from the fact that $\sup_{\o\in\R} |H(i\o)|<+\infty$ for  $H=\F h$.
In addition, if $x\in \U(r)$, then $y/\sup_{\o\in\R} |H(i\o)|\in \U(r)$.
In general, the process $y$ represents certain smoothing  of $x$, since  $\sup_{\o\in\R}|\o|^k  H(i\o)<+\infty$ for all $k\in \R$; however,  $h\notin \X(r)$.  \comm{It can be also noted that if $y\in \X(r)$
then there exists $\e\in (0,r)$ such that if $x\in \X(r-\e)$. }
\end{remark}
\begin{remark}
In Dokuchaev (2010), \index{\cite{D10},} a related linear weak predictability
 was established for processes from $\X(r)$. However, 
 the predictability  therein was established for 
a predictor  requiring infinite history 
of observations.  Theorem \ref{ThM} above establishes predictability
 with predictors requiring  a finite period of historical observations.
\end{remark}

\subsection{A family of predictors}

The question arises how to find the predicting kernels. We suggest
a possible choice of the kernels; they are given explicitly in the
frequency domain.
\par
 Let  $h\in\H(-\t,T)$ and $H=\F h$.\par

By the choice of $h$, we have that $H\ew=Q\ew e^{i\o T}$, where $Q\in \HH^2\cap \HH^\infty$
is such that $Q=\L q$, where $q(t)\equiv h(t-T)$. 

Let $H\ew|_{\o\in\R}$ be extended on $\C$ as $H(z)=Q(z)e^{ zT}$, $z\in\C$.
 
\vspace{0.4cm}   
\begin{theorem}\label{Th2} The following holds.
\begin{itemize}
\item[(i)]
There exists a
sequence of  polynomials  $\{\psi_d(z)\}_{d=1}^\infty$  of order $d$ such that 
\baa
&&\|e^{iT\cdot}-\psi_d(i\cdot)\|_{L_{2,-r}(\R,\C)}^2\breakk=\int_{-\infty}^\infty |e^{iT\o}-\psi_d(i\o)|^2  e^{-r|\o|}d\o\to 0\quad
\breakk \hbox{as}\quad d\to +\infty. \qquad
\eaa 
\item[(ii)] 
For $d=1,2,....$, $z\in\C$, set 
\baaa
\w h_d(t)\defi \sum_{k=0}^d a_{dk}\frac{d^kh}{dt^k}(t+T),
\eaaa
where $a_{dk}$ are the coefficients of the polynomials $\psi_d(z)=\sum_{k=0}^d a_{dk} z^k$.
Then  \comm{\OLD$\w h_d(\cdot)\in {\w H(T+\tau)}????$} $\w h_d(\cdot)\in\w\H(0,T+\t)$ for all $d$,  and the predictability
of the processes considered in Theorem \ref{ThM}(i)-(ii) can be ensured with
 the sequence of these  predicting  kernels, i.e., with
\baa
\w y_d(t)=\int^t_{t-T-\t}\w h_d(t-s)x(s)ds.\label{TD}
\eaa
\end{itemize}
\end{theorem}
\short{\vspace{0.4cm}  }
\par
Theorem \ref{Th2} describes predictor kernels in the time domain. 
These predictors can be represented explicitly in the frequency domain via their transfer functions 
\baa \w H_d(z)
\defi e^{-Tz} \psi_d (z) H(z),\quad
 \w h_d=\F^{-1}\w H_d|_{i\R}.\hphantom{}\label{Kk}
\label{FD} \eaa

Clearly, if the coefficients of a polynomial $\psi_d$ are real, then the predicting kernel $h_d(t)$ is real valued. In any case, if the underlying process $x$ is real valued, then one should replace $\w h_d(t)$ by its real part.

\short{In the next section, some choices of the polynomials  will be suggested.}

\section{On  selection of polynomials $\psi_d$}\label{secPol}
\short{A sequence of} Polynomials $\{\psi_k\}_{k=0}^{\infty}$ required for the predictors \short{described above } can be constructed from projections of the function  $e^{i\o T}$ on the 
truncated orthonormal basis in the Hilbert space $L_{2,-r}(\R,+\infty)$ using  the Gram--Schmidt 
procedure as the following.

Let $u_k(\o)\defi \o^k$,  $k=0,1,2,...$, and let 
\baaa
&&w_0=v_0=u_0/\|u_0\|_{L_{2,-r}(\R,\C)},\\&& v_k= u_k-\sum_{p=0}^{k-1}\frac{(u_k,v_p)_{L_{2,-r}(\R,C)}}{ \|v_p\|^2_{L_{2,-r}(\R,C)}}v_p,\quad \breakk w_k=v_k/\|v_k\|_{L_{2,-r}(\R,\C)},\quad k=1,2,...\, .
\eaaa
In this case,   \baaa
&&\|w_k\|_{L_{2,-r}(\R,C)}=1,\quad k=0,1,...,\qquad\breakk
(w_k,w_l)_{L_{2,-r}(\R,C)}=0,\quad  k,l=0,1,...,\quad k\neq l.
\eaaa
\begin{theorem}\label{ThGS} Let $c_k\defi (e^{iT\cdot},w_k)_{L_{2,-r}(\R,C)}$, $k=0,1,2..$.
Then the polynomials
\baaa
\psi_d^{\GS}(\o)\defi \sum_{k=0}^d c_k w_k(\o) \eaaa
are such as required in Theorem \ref{Th2}(i). Moreover, they are optimal in the sense that
\baaa
\|e^{iT\cdot}-\psi^{\GS}_d(i\cdot)\|_{L_{2,-r}(\R,\C)}\le \|e^{iT\cdot}-\psi_d(i\cdot)\|_{L_{2,-r}(\R,\C)}
\eaaa
for any $d$ and any polynomial $\psi_d$ of order d.
\end{theorem}
\par
For the case of small  prediction horizon $T<r$, the polynomials
$\psi_d$  can be constructed explicitly (although optimality in the sense of Theorem \ref{ThGS}(iii) will not be preserved).  
\short{\vspace{0.4cm}  }
\begin{theorem}\label{ThPol} 
For the case where $T<r$, the polynomials 
 $\psi_d(z)$ satisfying the assumptions of Theorem \ref{Th2}(i)  can be constructed   as $\psi_{d}(z)=\sum_{k=0}^d \frac{T^k z^k}{k!}$, i.e., as truncated  Taylor expansions of $e^{Tz}$.
 \end{theorem}

It can be noted that the choice of polynomials  in Theorems \ref{ThGS}-\ref{ThPol} 
 depends only on $T$ and $r$ only.

\vspace{0.4cm}  
\section{On robustness with respect to noise
contamination}\label{secRob} It is shown below that the predictors introduced in Theorem \ref{Th2} and
designed for processes from $\X(r)$ feature some robustness
with respect to noise contamination.   

Suppose that $r>0$ and either $p=1$ or $p=2$ is given.

 Assume that the  predictors  are
applied to a process $x\in L_2(\R,C)$ 
such that $x=x_0+\eta$, where $x_0\in\X(r)$, and where
$\eta\in L_p(\R,C)\cap L_2(\R,\C)$ represents the
noise. We assume that  either $p=1$ or $p=2$. 

Let $X=\F x$, $X_0=\F x_0$, and $N=\F \eta$.

We
assume that
$X_0(i\cdot)\in L_2(\R,C)$  and $\|N(i\cdot)\|_{L_p(\R,\C)}=\nu$.   The parameter $\nu\ge 0$ represents the
intensity of the noise.
\par
By the assumptions,  the predictors  are constructed as in Theorem \ref{Th2}
under the hypothesis that $\nu=0$, i.e. that 
$x=x_0\in\X$. By Theorems \ref{ThM}-\ref{Th2}, for an arbitrarily small $\e>0$, there exists $d$ such
that, if the hypothesis that $\nu=0$ is correct, then
\baaa &&E_d\defi \|\w
y_{d,0}-y\|_{L_{\infty}(\R,\C)}\le \e,\label{eps}\eaaa
where $\w y_{d,0}=\w h_{d} * x$ be defined via convolutions as in Theorem \ref{Th2} with $\nu=0$.

Let us estimate the prediction error for the case where $\nu>0$.
Let $\w y_{d,\eta}$ be defined by (\ref{TD}) with $x=x_0+\eta$. 
 We
have that \baaa \|\w y_{d,\eta}-y\|_{L_{\infty}(\R,\C)} \le E_d+ E_{\eta,d},\eaaa
where\baaa 
&&E_{\eta,d}= \|\w h_d *\eta-h* \eta\|_{L_{\infty}(\R,\C)}
\eaaa
represents the additional error caused by the presence of a high-frequency noise $\eta\notin\X(r)$  (when $\nu>0$).
We have that 
\baaa E_{\eta,d}\le \frac{1}{2\pi}\|(\w H_d(i\cdot)-H(i\cdot))N(i\cdot)\|_{L_1(\R,\C)}.
\eaaa It
follows that \baa&& \|\w y-y\|_{L_{\infty}(\R,\C)}\breakk
\le
\e+\frac{\nu}{2\pi}( \|\w H_d(i\cdot)\|_{L_q(\R,\C)}+\|H(i\cdot)\|_{L_q(\R,\C)}),\hphantom{xxx}\label{yn}\eaa
 where $q=+\infty$ for $p=1$ and $q=2$ for $p=2$.
\par
Therefore, it can be concluded that the prediction  is robust with
respect to noise contamination for any given $\e$. On the other
hand, if $\e\to 0$ then $\g\to +\infty$ and $\varkappa\to +\infty$. In
this case, the right hand part of  (\ref{yn}) is increasing for any given $\nu>0$.
Therefore, the error in the presence of noise will be large for a predictor targeting too small a size of the
error for the noiseless  processes from $\X(r)$.
\par
The equations describing the dependence of $(\e,\w\varkappa_d)$ on $d$
could be derived similarly to estimates in Dokuchaev (2012), \index{ \cite{D12},} Section 6,
where discrete time setting was considered. We leave it for future research.
\rd{\section{On numerical implementation}
\label{SecN}
\subsection{An algorithm based on time discretisation}

The predictor described above requires to calculate higher order derivative of the kernel $h$. 
The property of the admissible  kernel $h$ make it difficult to find its derivatives even in rare cases
where these derivatives can be found explicitly. For example, this  is the case for the kernel
\baa
h(t)=\frac{2}{T}\kappa_1\left(\frac{2(t-T/2)}{T}\right).
\label{exh} \eaa
that belongs to the class $\H(-T,0)$. 

We suggest to streamline calculations via replacing  the derivatives fro $h$ by the corresponding finite differences. 

Let us consider the problem of forecasting of the integral  \baa
A=\int_0^T h(T-s)x(s)ds
\label{A}\eaa
based on observations of $x(t)$ for $t\in [-T,0]$, where $h$ is defined by (\ref{exh}). 
This case is covered by Theorem \ref{ThM} with $\t=0$.

Let us select an integer $n>0$; this will be the number of sampling points $t\in [-T,0]$
for the observable process $x(t)$. Let $\vec{t}=\{t_k\}_{k=1}^n\in\R^n$ 
be such that  $0=t_1<t_2<\cdots <t_n=T$. We will use observations  $\{x(t_k-T)\}_{k=1}^n$.

For a function $f:\R^n\to  \R$ be a function, we define a vector $D(\vec{t},f)\in\R^n$
such that its $k$th component is $(f(t_{k+1})-f(t_k))/(t_{k+1}-t_k)$ for $k=1,...,n-1$,
and that its $n$th component is zero. \comm{ the same as its $(n-1)$th component.} 

It can be noted that the first $n-1$ components of the vector $D(\vec{t},f)$ are finite differences approximating the derivative $df/dt$  as $n\to +\infty$ for differentiable functions $f$.  We select
the last component to be zero since we will need to calculate this vector only for functions $f$
vanishing at the endpoints together with all derivatives; in that case, zero is the best approximation. 

Let an integer $d>0$ be given, and let $a_{dk}$ be defined as in Theorem \ref{ThM}.

For  $h\in\H(-T,0)$, let $q(t)=h(t+T)$, and let
\baaa
\D_0  =\{q_j\}_{j=1}^n,\quad \D_1 =D(\vec{t},\D_0),...,\D_k =D(\vec{t},\D_{k-1}),....
\eaaa
\def\h{\w{\rm h}}
\def\EE{{\mathbb{E}}}

Let $\h\in\R^n$ be defined as\baaa
\h=\sum_{k=0}^d a_{kd} \D_k. 
\eaaa 

Then the estimate $
\w A_{n,d}$  of $A$
 can be calculated as
 \baa
 \label{wA}
 \w A_{n,d}=\sum_{k=1}^{n-1} \h(n-k)x(t_k-T) (t_{k+1}-t_k).
 \eaa
This is an approximation of estimate (\ref{TD}) after discretization in time.  
\subsection{Some numerical experiments}

We made some numerical experiments.

For selected varying $T>0$ and $d>0$,
 
 We considered  input  processes $x\in\X(r)$ with $r>T$ obtained via the Monte-Carlo simulation
 as the following.
 \begin{enumerate}
 \item We calculated vectors $\vec{h}$
using  coefficients $a_{dk}=T^k/k!$ such as described in Theorem \ref{ThPol}; by the choice of $y$, we have
 \item We simulated random input $x(t)$. At each Monte-Carlo simulation, be selected independent random numbers $a,b,c,d,e,f$ 
 uniformly distributed over intervals $[-500,500]$, 
 independent random numbers $p,q$  uniformly distributed over intervals $[1,20]$
  and 
 independent random numbers $\a,\b$  uniformly distributed over intervals $[0.05,500]$.
 For each set of number, we defined a process 
 \baaa
 \oo x(t)=\frac{a}{p+T+iT}+\frac{b}{q+T+iT}+c\,\sinc (t/\a+e) +d\,\sinc (t/\b+f) 
 \eaaa
 It can be noted that the first two terms in this sum represent processes from $\X(p+T)$ and $\X(q+T)$ respectively, and the second two terms represent band limited processes
 that belong to $\X(r)$ for any $r>0$.  This means   that 
 $\oo x\in \X(r)$ for some $r>T$.
\item  For some given $\s\ge 0$,  random noise  processes  $\eta$ were simulated
  as Gaussian process $\{\eta(t)\}_{t=-n}^n$ with independent
values such that $\E \eta_r(t)=0$ and $\Var \eta_r(t)=\s^2$.
 A noise contaminated process  $x=\oo x+\eta_r$ was created to replace $\oo x$ in the simulation.
  \item For each $x(\cdot)$, we calculated the value $A$ defined by (\ref{A}).
  \item For each $x(\cdot)$, we calculated the value $\w A_{n,d}$ defined by (\ref{wA}).
 \item For each $x(\cdot)$, we calculated the forecast error $\w A_{n,d}-A$ and the relative
 forecast error 
 \baaa
 {\cal E}(n,d,\sigma,x(\cdot))=\frac{|\w A_{n,d}-A|}{|A|}
 \eaaa
 \item 
We calculated  the mean relative error
\baaa
E(n,d,\sigma)=\EE {\cal}  {\cal E}(n,d,\sigma,x(\cdot)).
\eaaa
Here $\EE$ means the average over the Monte-Carlo simulations.
 \end{enumerate}

We have used R software. 
 We have used $T=0.2$, and we have used 1000 Monte-Carlo simulations
 for each set of parameters.

For the case where $\sigma=0$, i.e., without the additional noise, we obtained  that
 \begin{itemize}
 \item[]
 $E(500,5,0)=0.026$,\quad\quad  $E(1000,5,0)=3\cdot 10^{-6}$,
 \item[]
$E(3000,4,0)=7\cdot 10^{-6}$, \quad $E(3000,5,0)=1\cdot 10^{-6}$.
\end{itemize}
\par
It can be noted that even in the case where $\s=0$ (i.e., where the random noise $\eta_r$ is absent), the forecasting algorithm described in Theorem \ref{ThM}  
was actually applied to discrete time  approximations of processes from $\X(r)$, which corresponds to the presence of some noise generated by the discretization error.

In addition, we made some experiment on the data recovery in the presence of noise.
We obtained  that
\begin{itemize}
 \item[]
$E(20000,4,0.05)=0.121$, \quad  $E(40000,4,0.05)=0.069$,
\item[]
$E(4000,4,0.02)=0.106$, \quad   $E(3000,4,0.02)=0.1265$,
 \item[]
$E(2500,4,0.01)=0.066$, \quad $E(3000,4,0.01)=0.060$,
\item[]
$E(4000,4,0.01)=0.051$, \quad $E(10000,4,0.01)=0.026$.
\item[]
\end{itemize}

The results of these experiments confirm   the effectiveness of the method and some
robustness with respect to noise contamination.
}
 \section{Proofs}\label{secProof}
Theorem \ref{ThM} follows immediately from Theorem \ref{Th2}. \short{ Let us prove Theorem \ref{Th2}.}
\par
{\em Proof of Theorem \ref{Th2}}. By  the Completeness Theorem for polynomials  (Higgins (1977), p.31),
 \index{\cite {Higgins}, p.31,} it follows 
 that  there exists 
sequence of  polynomials  $\{\oo\psi_d(\o)\}_{d=1}^\infty$ in $\o\in \R$ of order $d$ such that 
\baaa
&&\|e^{iT\cdot}-\oo\psi_d(\cdot)\|_{L_{2,-r}(\R,\C)}^2\breakk =\int_{-\infty}^\infty |e^{iT\o}-\oo\psi_d(\o)|^2  e^{-r|\o|}d\o\to 0\quad \short{\breakk }\hbox{as}\quad d\to +\infty. 
\eaaa
The coefficients $a_k$ of desired polynomials $\psi_d(z)=\sum_{k=0}^d a_k z^k$ can be constructed by adjustment the signs of the coefficients for the polynomials
$\oo\psi_d(\o)=\sum_{k=0}^d \oo a_k \o^k$ such that $\psi_d(i\o)\equiv \oo\psi_d(\o)$, i.e., $\oo a_k=a_k i^k$
and $a_k=\oo a_k i^{-k}$. This proves statement (i).

Let us prove statement (ii).  Clearly, $q(t)=0$ for $t<0$ and $q\in C^{\infty}(\R)$. Hence $z^n Q(z)\in \HH^2\cap\HH^\infty$ for any integer $n\ge 0$.
It follows  that \baa
\w H(z)=\psi_d(z)Q(z)\in \HH^2\cap\HH^\infty.
\label{1}\eaa
Further, $Q(z) e^{(T+\t)z}\in \HH^2_-$.
Hence
 \baa
&&\w H(z)e^{(T+\t)z}=e^{-T z}\psi_d(z) e^{Tz}Q(z) e^{(T+\t)z}\breakk =\psi_d(z)Q(z) e^{(T+\t)z} \in \HH^2_-.
\label{2}\eaa
It follows from (\ref{1})-(\ref{2}) that $\w H\in \H(-\t,T)$.

\par
 For  $x\in\X(r)$,
let
 $X(i\o)= \F x$,
 $Y(i\o)= \F y=H(i\o)X(i\o)$, $\w Y_d(i\o)= \w H_d(i\o)X(i\o)$, and $\w y=\F^{-1}\w Y_d$.
\par
We have that
\baa
&&\|\w y_{d}-y\|_{L_\infty(\R)}\breakk \le \frac{1}{2\pi} \|(\w H_d(i\cdot) -H(i\cdot))X(i\cdot)\|_{L_1(\R)}.
\label{est1}\eaa
Furthermore, 
\baa
&&\|(\w H_d(i\cdot) -H(i\cdot))X(i\cdot)\|_{L_1(\R)}
\breakk=\int_{-\infty}^\infty \Bigl|(e^{- i\o T}\psi_d(i\o)-1) 
e^{ i\o T} Q(i\o)X(i\o)\Bigr|d\o
\nonumber \\
&&= \int_{-\infty}^\infty e^{-r|\o|/2}\Bigl| (e^{- i\o T}\psi_d(i\o)-1) \BRR
e^{r|\o|/2}e^{ i\o T} Q(i\o)X(i\o)\Bigr|d\o 
\nonumber\\
&&=\int_{-\infty}^\infty e^{-r|\o|/2}\Bigl|(\psi_d(i\o) -e^{i\o T}) 
\short{\BRR}
\hphantom{x}
\BRR 
e^{r|\o|/2}e^{ i\o T} Q(i\o)X(i\o)\Bigr|d\o 
\short{\breakk}\le \a_d^{1/2}  \b^{1/2}. \label{est}
\eaa
Here
\baaa
&&\a_d=\int_{-\infty}^\infty e^{-r|\o|}|\psi_d(i\o)-e^{i\o T}|^2d\o,\qquad 
\breakk \b=\int_{-\infty}^\infty e^{r|\o|}|e^{ i\o T} Q(i\o)X(i\o)|^2d\o.
\eaaa
By the choice  of $\psi_d$, it follows that
\baa
 \a_d\to 0\quad \hbox{ as}\quad d\to +\infty.
\label{a} 
\eaa
Since $q\in C^\infty(\R;\R)$ and has a bounded support, it follows that 
 $\sup_\o |Q(i\o)|<+\infty$.   Hence
\baa
|\b|\le \sup_\o |Q(i\o)| \int_{-\infty}^\infty e^{r|\o|} |X(i\o)|^2d\o.
\label{b}\eaa
We have that $\b$ is finite for each $X$  under the assumptions of of Theorem \ref{Th2}(i)-(ii), and
that $\b$ is bounded over $x\in\U(r)$  under the assumptions of of Theorem \ref{Th2}(ii). Then estimates (\ref{est1})--(\ref{b}) imply the proof of Theorem \ref{Th2}. $\Box$.

{\em Proof of Theorem \ref{ThGS}} follows from the completeness
of polynomials in $L_{2,-r}(\R,\C)$ (Higgins (1996), p.31),
\index{(\cite{Higgins}, p.31),} and from 
the orthonormality of the sequence $\{w_k\}$ implied by the properties of the   
Gram-Schmidt orthogonalization process. 

{\em Proof of Theorem \ref{ThPol}.} 
 We have that 
 \baaa \psi_{d}(i\o)=\sum_{k=0}^d \frac{(Ti \o)^k}{k!}=C_d(\o)+iS_d(\o),
 \eaaa
where
\baaa
&& C_{d}(i\o)=\sum_{m=0}^\infty \frac{(Ti\o)^{2m}}{(2m)!}= (-1)^{m}\sum_{m=0}^\infty \frac{T^{2m} \o^{2m}}{(2m)!} \eaaa
and
\baaa
 S_{d}(i\o)=\frac{1}{i}\sum_{m=0}^\infty \frac{(Ti\o)^{2m+1}}{(2m)!}=(-1)^{m}\sum_{m=0}^\infty \frac{T^{2m+1}  \o^{2m}}{(2m)!}. 
\eaaa
We have that $C_d(i\o)$ and $S_d(i\o)$ are truncated Taylor  expansions for $\cos(T\o)$ and $\sin(T\o)$ respectively.
   Hence   
\baaa
&&\int_{-\infty}^\infty |\cos(T\o)-C_{d-1}(i\o)|^2  e^{-r |\o|}d\o\breakk \le \int_{-\infty}^\infty \frac{T^d}{d!} |\o|^d e^{-r |\o|} d\o
=
2\frac{T^d}{d!} \frac{d!}{r^{d-1}}= \frac{2T^{d}}{r^{d-1}}.
\label{L10}\eaaa
By the assumptions,  we have that $T<r$.  Hence
\baaa
\int_{-\infty}^\infty |\cos(T\o)-C_{d}(i\o)|^2  e^{-r| \o|}d\o\to 0 \quad\hbox{as}\quad d\to+\infty. \hphantom{xxxx} 
\label{L1}\eaaa
Similarly, we obtain that
\baaa
\int_{-\infty}^\infty |\sin(T\o)-S_{d}(i\o)|^2  e^{-r| \o|}d\o\to 0\quad\hbox{as}\quad d\to+\infty.  
\label{L2}\eaaa
This completes the proof of Theorem \ref{ThPol}. $\Box$
\short{\begin{remark}
Let $r>0$ and $q>0$. Let  $L_{2,r,q}(\R;\C)$ be the Hilbert space of complex valued functions $u:\R\to\C$ with the norm 
 \baaa \|u\|_{L_{2,r,q}(\R,\C)}=\Bigl(\int_{-\infty}^{+\infty}e^{r|\o|^q}|u(\omega)|^2 d\o\Bigr)^{1/2}.\eaaa
 It can be shown  that the proof implies that the polynomials cannot form a complete set in $L_{2,-r,q}(\R;\C)$ for any $r>0$ and $q\in (0,1)$. Otherwise, the direct  implementation 
 of the proof of Theorem \ref{ThM}  would imply predicability in the sense of Definition \ref{def1}
 of non-zero processes $x\in L^2(\R)$ such that $x(t)=0$ for $t<0$, which is impossible.
\end{remark}
}
 \section{Discussion and future research}\label{secCon}
The present paper studies  prediction of  continuous time processes in pathwise deterministic setting. The  paper suggests  
linear integral predictors with limited memory for prediction of  anti-causal  convolutions
with finite horizon.  

The predictors are  described explicitly via polynomials approximating periodic exponents (complex sinusoids)
defined by the predicting horizon.
\begin{enumerate}
 \item 
 The predictors do not depend on the shape of the spectrum of the underlying process.  
 \item
The predictors   are not error-free; however, the error can be made
arbitrarily small with a choice of larger degrees of polynomials. 
\item
Some predictors  for the same class of underlying processes were 
obtained earlier in  Dokuchaev (2010). However, the predictors in  were quite different: they required unlimited history of observations. \index{\cite{D10}}
\rd{\item
The method leads to a relatively simple numerical  algorithm based on time discretization. 
\item
Predictors feature some robustness with respect to noise contamination. This means that the predictors can be applied for processes that are not necessarily  in the class $\X(r)$, provided that a certain forecasting error is tolerable.  }  
\end{enumerate}

Since  $\sup_\o |\w h_{d}(i\o)|$ could increase  fast as $d\to +\infty$, 
the method  would  require calculations with  large numbers
to achieve high predicting accuracy.  We leave this for the future research. 


\section*{References}
 \
 
Dokuchaev, N.G. (2008). The predictability of band-limited,
high-frequency, and mixed processes in the presence of ideal
low-pass filters.   {\it Journal of Physics A: Mathematical and
Theoretical} {\bf 41} No 38, 382002 (7pp)

   Dokuchaev, N. (2010). Predictability on finite horizon
for processes with exponential decrease of energy on higher
frequencies.
 {\it Signal Processing} {\bf 90} Iss. 2, 696--701. 

Dokuchaev, N. (2012). Predictors for discrete time processes with
energy decay on higher frequencies. {\em IEEE Transactions on Signal
Processing} {\bf 60}, No. 11, 6027-6030.

Dokuchaev, N. (2021). Pathwise continuous time weak predictability and single point spectrum degeneracy. 
{\em  Applied and Computational Harmonic Analysis}, (53) 116–131.
\comm{ \bibitem{D17}  Dokuchaev, N. (2017).
On  linear weak predictability with single point spectrum degeneracy. arXiv:1705.02746.}

Duren P. (1970) {\it Theory of $H^p$-Spaces.}  Academic Press, New
York.

Higgins, J.R. (1977).
 Completeness and Basis Properties of Sets of Special Functions.
Cambridge University Press. 1977.

Higgins, J.R. (1996). Sampling Theory in Fourier and Signal
Analysis. Oxford University Press, New York.

Knab J.J. (1979). Interpolation of band-limited functions using
the approximate prolate series. {\it IEEE Transactions on
Information Theory} {\bf 25}(6), 717--720.

Lyman R.J, Edmonson W.W., McCullough S., and Rao M. (2000). The
predictability of continuous-time, bandlimited processes. {\it
IEEE Transactions on Signal Processing} {\bf 48}, Iss. 2,
311--316.

Lyman R.J and Edmonson W.W. (2001). Linear prediction of
bandlimited processes with flat spectral densities. {\it IEEE
Transactions on Signal Processing} {\bf 49}, Iss. 7, 1564--1569.

Marvasti F. (1986). Comments on "A note on the predictability of
band-limited processes." {\it Proceedings of the IEEE}, {\bf
74}(11), 1596.

Papoulis A. (1985). A note on the predictability of band-limited
processes. {\it Proceedings of the IEEE}, {\bf 73}(8), 1332--1333.

Slepian D. (1978). Prolate spheroidal wave functions, Fourier
analysis, and uncertainty--V: The discrete case. {\it Bell System
Technical Journal}, {\bf 57}(5), 1371--1430.

 Vaidyanathan P.P. (1987). On predicting a band-limited signal
based on past sample values. {\it Proceedings of the IEEE}, {\bf
75}(8), 1125--1127.


\comm{

}
\end{document}